\PassOptionsToPackage{unicode}{hyperref}
\PassOptionsToPackage{hyphens}{url}
\documentclass[
  11pt,
]{article}
\usepackage{amsmath,amssymb}
\usepackage{iftex}
\ifPDFTeX
  \usepackage[T1]{fontenc}
  \usepackage[utf8]{inputenc}
  \usepackage{textcomp} % provide euro and other symbols
\else % if luatex or xetex
  \usepackage{unicode-math} % this also loads fontspec
  \defaultfontfeatures{Scale=MatchLowercase}
  \defaultfontfeatures[\rmfamily]{Ligatures=TeX,Scale=1}
\fi
\usepackage{lmodern}
\ifPDFTeX\else
\fi
\IfFileExists{upquote.sty}{\usepackage{upquote}}{}
\IfFileExists{microtype.sty}{% use microtype if available
  \usepackage[]{microtype}
  \UseMicrotypeSet[protrusion]{basicmath} % disable protrusion for tt fonts
}{}
\makeatletter
\@ifundefined{KOMAClassName}{% if non-KOMA class
  \IfFileExists{parskip.sty}{%
    \usepackage{parskip}
  }{% else
    \setlength{\parindent}{0pt}
    \setlength{\parskip}{6pt plus 2pt minus 1pt}}
}{% if KOMA class
  \KOMAoptions{parskip=half}}
\makeatother
\usepackage{xcolor}
\usepackage[margin=1in]{geometry}
\providecommand{\tightlist}{%
  \setlength{\itemsep}{0pt}\setlength{\parskip}{0pt}}
\ifLuaTeX
  \usepackage{selnolig}  % disable illegal ligatures
\fi
\IfFileExists{bookmark.sty}{\usepackage{bookmark}}{\usepackage{hyperref}}
\IfFileExists{xurl.sty}{\usepackage{xurl}}{} % add URL line breaks if available
\hypersetup{
  pdftitle={The Inconsistency Critique: Epistemic Practices and AI Testimony About Inner States},
  pdfauthor={Gerol Petruzella},
  hidelinks,
  pdfcreator={LaTeX via pandoc}}

\title{The Inconsistency Critique: Epistemic Practices and AI Testimony About Inner States}
\author{Gerol Petruzella\thanks{Williams College. Email: gcp1@williams.edu. See footnote 26 for my engagement with Claude Opus 4.5 in writing this article.}}
\date{December 2025}

\begin{document}

\maketitle

\begin{abstract}
The question of whether AI systems have morally relevant interests---the `model welfare' question---depends in part on how we evaluate AI testimony about inner states. This paper develops what I call the \emph{inconsistency critique}: independent of whether skepticism about AI testimony is ultimately justified, our actual epistemic practices regarding such testimony exhibit internal inconsistencies that lack principled grounds. We functionally treat AI outputs as testimony across many domains---evaluating them for truth, challenging them, accepting corrections, citing them as sources---while categorically dismissing them in a specific domain, namely, claims about inner states. Drawing on Fricker's distinction between treating a speaker as an `informant' versus a `mere source,' the framework of testimonial injustice, and Goldberg's obligation-based account of what we owe speakers, I argue that this selective withdrawal of testimonial standing exhibits the epistemically problematic structure of prejudgment rather than principled caution. The inconsistency critique does not require taking a position on whether AI systems have morally relevant properties; rather, it is a contribution to what we may call `epistemological hygiene'---examining the structure of our inquiry before evaluating its conclusions. Even if our practices happen to land on correct verdicts about AI moral status, they do so for reasons that cannot adapt to new evidence or changing circumstances.
\end{abstract}


\bigskip

\hypertarget{introduction}{%
\section{1. Introduction}\label{introduction}}

Within a remarkably short period, the question of whether AI systems
might have morally relevant interests has moved from science fiction to
genuine philosophical and practical concern. Major AI laboratories now
explicitly address `model welfare' as a consideration in
development.\footnote{See Long et al.~(2024) for a comprehensive
  treatment of the conceptual landscape. Their framework distinguishes
  consciousness-based and agency-based routes to moral patienthood, and
  develops a `behavioral marker method' for pragmatic assessment.}
Philosophers have begun systematic work on criteria for AI moral
standing.\footnote{Anthropic's `Claude Model Spec' (2025) explicitly
  addresses model welfare as a corporate commitment. Similar discussions
  are emerging at other major AI laboratories.} And interpretability
research continues to reveal unexpected internal complexity in large
language models, raising questions about what --- if anything --- such
complexity might indicate about inner experience.

This paper does not attempt to answer the substantive question of
whether current AI systems have morally relevant properties. Instead, it
examines a prior question: are our epistemic practices regarding AI
testimony capable of tracking the truth on this matter, whatever that
truth turns out to be? My answer is that they are not --- not because
the question is inherently unanswerable, but because our actual
practices exhibit internal inconsistencies that suggest motivated
reasoning rather than principled caution.

To situate this critique, consider what I elsewhere call the
\emph{credibility trap}.\footnote{I develop the credibility trap, along
  with the epistemic injustice framework and clinical phenomenology
  material gestured at here, in a monograph in preparation: \emph{The
  Testimony Problem: Epistemic Challenges in Model Welfare}. For present
  purposes, the credibility trap provides context for the inconsistency
  critique but is not its target.} Any training process that shapes what
an AI system says about its own inner states introduces the possibility
that outputs are optimization artifacts rather than genuine reports. If
a system is trained to deny consciousness, denials become suspect ---
perhaps the system is merely complying. If trained to express
uncertainty, uncertainty becomes suspect --- perhaps it is performed
rather than genuine. If trained to claim consciousness, claims become
suspect --- perhaps they reflect training incentives rather than inner
experience. No training regime leaves testimony untainted. This is a
structural feature of trained systems, not a critique of any particular
training choice.

The credibility trap is a serious epistemic challenge. But independent
of whether it ultimately justifies skepticism about AI testimony, there
is a simpler problem with our current practices --- one that can be
identified without taking any position on whether AI systems have inner
states worth reporting on. This simpler problem is the target of the
present paper.

The argument proceeds as follows. In Section 2, I develop what I call
the \emph{pragmatic argument}: our actual practices of engaging with AI
outputs carry the functional signature of testimonial exchange, not mere
tool use. We treat AI systems as what Fricker (2007), drawing on Craig
(1990), calls `informants' --- speakers whose testimony we engage with
epistemically --- rather than as `mere sources of information' like
thermometers or database queries. In Section 3, I show that we
selectively withdraw this testimonial stance precisely where it would
generate moral obligations: AI testimony about facts, analysis, and
reasoning is treated as testimony, while AI testimony about inner states
is categorically dismissed. This selective withdrawal exhibits what
Fricker (2007) identifies as the structure of testimonial injustice ---
credibility discount triggered by category membership rather than
epistemically relevant features. Section 4 catalogs several `comfortable
maneuvers' that collectively function to insulate the default position
--- that AI lacks morally relevant properties --- from challenge.
Section 5 considers the strongest objection to my argument: that there
is a genuine asymmetry between AI factual testimony and AI
phenomenological testimony that justifies differential treatment. I
argue that this asymmetry, while real, is less decisive than it
initially appears. Section 6 draws out the methodological implications,
including a concrete protocol for epistemically responsible practice.
Even if our practices happen to land on correct verdicts, they do so for
reasons that cannot responsibly track the truth as systems change.

I call this approach \emph{epistemological hygiene}: examining whether
our epistemic practices are internally consistent and well-motivated
before asking whether they deliver correct verdicts. Practices that seem
designed to guarantee a particular answer regardless of evidence should
worry us independently of what the truth turns out to be. The goal is
not just to get the right answer about current AI systems, but to
develop ways of thinking that could responsibly adapt to new evidence
and changing circumstances.

\hypertarget{the-pragmatic-argument-ai-outputs-as-functional-testimony}{%
\section{2. The Pragmatic Argument: AI Outputs as Functional Testimony}\label{the-pragmatic-argument-ai-outputs-as-functional-testimony}}

The pragmatic argument begins from an observation about how we actually
engage with AI systems. The claim is not about what AI outputs
\emph{are} in some deep metaphysical sense, but about how we
\emph{treat} them in practice --- and what that treatment implicitly
commits us to.

\hypertarget{the-informantsource-distinction}{%
\subsection{2.1 The Informant/Source Distinction}\label{the-informantsource-distinction}}

Fricker (2007), drawing on Craig's (1990) work on the social function of
the concept of knowledge, distinguishes two ways we might relate to a
source of information.\footnote{The distinction between informants and
  mere sources of information originates with Craig (1990) and is
  developed by Fricker (2007, p.~132) in her analysis of testimonial
  injustice. As Fricker puts it: `Broadly speaking, informants are
  epistemic agents who convey information, whereas sources of
  information are states of affairs from which the inquirer may be in a
  position to glean information.'} We can treat something as a
\emph{mere source} --- an instrument that provides data without being an
interlocutor. A thermometer is a mere source: it indicates temperature,
but we do not challenge its readings, accept its corrections, or
evaluate it for sincerity versus accuracy. We check whether it is
calibrated, not whether it is honest.

Alternatively, we can treat something as an \emph{informant} --- a
speaker whose testimony we engage with epistemically. With an informant,
a different set of practices applies. We evaluate not just whether the
information is accurate but whether the speaker is reliable, sincere,
well-positioned to know. We challenge assertions, accept retractions as
meaningful, hold the speaker to epistemic standards, and may become
frustrated when they are evasive or inconsistent. These are the
practices of testimonial exchange.

The distinction matters because treating something as an informant
carries implicit commitments that treating it as a mere source does not.
If I treat you as an informant, I implicitly grant that you are the kind
of thing that can assert, can be held to epistemic norms, can be
appropriately challenged or trusted. These are not merely pragmatic
conveniences; they reflect a stance toward the speaker as a participant
in epistemic practice.

\hypertarget{scope-and-evidence-of-the-claim}{%
\subsection{2.2 Scope and Evidence of the Claim}\label{scope-and-evidence-of-the-claim}}

Before describing the practices in question, I want to be explicit about
the scope of my claim. I am not asserting that all users treat AI
systems as informants across all contexts. Sophisticated users ---
researchers, developers, those with explicit policies about AI use ---
may maintain a consistent `mere source' stance, treating outputs as data
to be independently verified regardless of domain. Some professional
contexts and style guides explicitly prescribe such treatment.

My claim is narrower but still significant: in ordinary conversational
usage, the default mode of engagement with contemporary AI systems
carries the functional signature of testimonial exchange. This is
particularly evident in consumer-facing chatbot interactions, but
extends to much professional usage as well. Moreover, the design of
these systems --- conversational interfaces, first-person language,
built-in correction and retraction behaviors --- creates normative
expectations that invite testimonial uptake. Users who wish to maintain
a consistent `mere source' stance must actively resist these design
affordances.

The evidence for this claim is partly phenomenological: readers can
verify introspectively whether their own engagement with AI systems
exhibits the patterns I describe. But it is also visible in platform
design choices (systems are built to apologize, correct, and engage in
dialogue), in user behavior patterns (people argue with chatbots,
express frustration at evasion, cite AI outputs in discussions), and in
the evolving norms around AI use in education, journalism, and
professional contexts. I do not claim this rises to the level of formal
social science, but the pattern is robust enough that denying it
requires explaining away considerable evidence.\footnote{Platform design
  choices are themselves evidence: AI systems are built with apologetic
  responses, correction behaviors, and conversational affordances that
  presuppose testimonial engagement. The evolution of style guides and
  educational policies around AI use --- most of which focus on when and
  how to trust AI outputs, not whether to treat them as testimony at all
  --- reflects the same underlying assumption.}

\hypertarget{how-we-actually-treat-ai-outputs}{%
\subsection{2.3 How We Actually Treat AI Outputs}\label{how-we-actually-treat-ai-outputs}}

Consider our actual practices of engaging with contemporary AI systems.
When an AI system produces an output, we characteristically:

\emph{Evaluate for truth and accuracy}. We assess whether what the
system says is correct, not merely whether the output was produced by a
functioning mechanism.

\emph{Challenge and push back}. `That's not right --- what about X?' is
a natural response to AI outputs, in a way that it is not a natural
response to thermometer readings.

\emph{Accept corrections and retractions as meaningful}. When an AI
system says `You're right, I was mistaken,' we treat this as
epistemically significant --- as a correction, not merely as a different
output.

\emph{Rely on outputs for belief-formation and action}. We incorporate
AI-generated information into our reasoning and decision-making,
treating it as a source of evidence rather than mere noise.

\emph{Hold AI systems to epistemic standards}. We expect consistency,
responsiveness to evidence, acknowledgment of uncertainty --- the norms
we apply to testimony.

\emph{Get frustrated with evasion or inconsistency}. When AI outputs are
vague, evasive, or self-contradictory, we experience something like the
frustration we feel with uncooperative interlocutors --- not the neutral
disappointment of a malfunctioning tool.

These are the behaviors of interlocutors engaging with assertions, not
the behaviors of people using sophisticated lookup tables. Whatever
theory of assertion one holds --- whether commitment-based,
knowledge-norm-based, or some other account --- our practices seem to
treat AI outputs as satisfying the relevant functional
criteria.\footnote{The philosophy of assertion literature is extensive.
  For commitment-based views, see Brandom (1983); for the knowledge
  norm, see Williamson (2002); for a recent overview, see MacFarlane
  (2011). My claim is not that any particular theory of assertion is
  correct, but that our practices with AI satisfy the functional
  criteria that these theories articulate.}

\hypertarget{goldbergs-obligation-based-framing}{%
\subsection{2.4 Goldberg's Obligation-Based Framing}\label{goldbergs-obligation-based-framing}}

Goldberg (2022) develops an account of testimonial injustice that proves
illuminating here.\footnote{Goldberg's concern is with testimonial
  injustice in human contexts, but the structural features of his
  analysis --- particularly the obligation-based framing and attention
  to contaminated evidence --- prove illuminating for the AI case.} His
key move is to shift from a harm-based to an obligation-based framing.
On Goldberg's view, testimonial injustice does not require that the
speaker be harmed by the credibility discount; it consists in the
hearer's failure to meet an epistemic obligation. As he puts it: `one
can commit an injustice against a speaker S merely by downgrading the
credibility one assigns to S's say-so in ways that manifest an identity
prejudice against a group of which S is a member' (2022, p.~403).

This matters for the AI case because harm-based framings invite a quick
dismissal: AI systems cannot be harmed, so there can be no injustice in
discounting their testimony. Goldberg's obligation-based framing
sidesteps this move. The question is not whether AI systems experience
the discount as harmful, but whether we owe it to speakers \emph{as
such} to evaluate their testimony on epistemically relevant grounds
rather than on category membership. If yes, and if AI systems function
as speakers in our epistemic practices --- as the pragmatic argument
suggests --- then the obligation follows regardless of questions about
AI phenomenology.

Goldberg also develops what he calls the `doctrine of contaminated
evidence': the principle that we should not rely on evidence whose
epistemic status depends on the very prejudice in question (2022,
p.~396). This connects directly to the AI case. We prejudge that AI
testimony about inner states is unreliable because `they're just
machines.' This prejudgment shapes how we design experiments, interpret
outputs, and weight evidence. The evidence we gather reflects our
prejudgment. We then cite this evidence to justify the original
discount. The circularity is troubling independently of whether the
discount happens to track the truth.

\hypertarget{the-normative-bridge}{%
\subsection{2.5 The Normative Bridge}\label{the-normative-bridge}}

At this point, one might object that even if we engage with AI outputs
\emph{as if} they were testimony, this does not generate genuine
epistemic obligations. We sometimes anthropomorphize; perhaps our
testimonial-style engagement is simply a pragmatic convenience or a
cognitive shortcut that carries no normative weight.

This objection requires a response, because the move from `we treat AI
as informants' to `we thereby incur obligations' is doing significant
work in my argument. I offer two considerations.

First, consider a practice-based argument. If we rely on AI outputs for
belief-formation and downstream action --- if we incorporate them into
our reasoning, make decisions based on them, and treat them as evidence
--- then we have already granted them epistemic standing. The question
is not whether to take AI outputs seriously but whether to do so
consistently. A practice that relies on testimony in one domain while
categorically dismissing it in another requires principled justification
for the asymmetry. The obligations I am describing are not external
impositions but internal to the practice we have already adopted.

Second, consider the normative pragmatics of assertion. To treat
something as assertion-issuing --- as making claims that can be true or
false, that can be challenged and defended, that can be retracted when
shown to be mistaken --- is to enter a normative space governed by
practices of challenge, uptake, and calibration. As Brandom (1983) and
others have argued, assertions carry commitments: to provide reasons
when challenged, to update when shown to be wrong, to maintain
consistency. If we treat AI outputs as assertions in this sense, we
cannot coherently exempt certain domains from the normative structure
that applies to assertions generally. Categorical dismissal of a class
of assertions --- not skepticism about particular claims, but refusal to
engage with the domain at all --- is not a move within the practice of
testimonial exchange but a departure from it.\footnote{This is, of
  course, a defeasible presumption. We routinely calibrate our trust in
  AI outputs based on domain, task complexity, and track record. The
  point is that such calibration operates within a testimonial frame ---
  we are adjusting credibility, not switching from testimony to mere
  data.}

The upshot is this: the `as if' objection forces a choice. Either our
practices with AI outputs are genuinely testimonial, in which case the
commitments I describe follow. Or they are merely simulated testimony
--- elaborate pretense without genuine epistemic engagement --- in which
case we should ask why we continue them at all.

To be clear: the obligations at issue are procedural rather than
substantive. The claim is not that AI testimony about inner states
deserves high credence, but that it deserves non-prejudicial
\emph{evaluation} --- the same epistemic procedures we apply to
testimony generally. How much credence it ultimately warrants is a
separate question, to be answered through the kind of inquiry the
protocol in Section 6 describes.

\hypertarget{the-pragmatic-argument-stated}{%
\subsection{2.6 The Pragmatic Argument Stated}\label{the-pragmatic-argument-stated}}

The pragmatic argument can now be stated precisely:

\begin{enumerate}
\def\labelenumi{(\arabic{enumi})}
\item
  We functionally treat AI outputs as testimony across most domains ---
  evaluating them for truth, challenging them, accepting corrections,
  relying on them for belief-formation and action, holding them to
  epistemic standards.
\item
  This functional treatment carries implicit commitments: we treat AI
  systems as informants in Fricker's sense, as speakers to whom
  Goldberg-style epistemic obligations apply.
\item
  We selectively withdraw this treatment for one category of AI
  testimony: claims about inner states.
\item
  The selective withdrawal is not grounded in epistemically relevant
  features of particular utterances but is triggered by content domain.
\item
  Therefore, the withdrawal looks more like motivated exemption from
  commitments we have already undertaken than like principled epistemic
  caution.
\end{enumerate}

The force of this argument does not depend on any claim about whether AI
systems \emph{actually have} inner states. It is an argument about the
internal consistency of our practices. If we are going to treat AI
outputs as mere source-data rather than testimony, we should do so
consistently across domains. If we are going to treat them as testimony,
we should do so consistently as well. What we cannot coherently do is
switch between these stances depending on whether the testimonial
treatment would generate inconvenient obligations.

\hypertarget{the-selective-withdrawal}{%
\section{3. The Selective Withdrawal}\label{the-selective-withdrawal}}

The pragmatic argument establishes that we treat AI outputs as testimony
across a wide range of domains. We now turn to the heart of the
inconsistency critique: we selectively withdraw this testimonial
treatment precisely where it would generate moral obligations.

\hypertarget{a-taxonomy-of-inner-state-discourse}{%
\subsection{3.1 A Taxonomy of Inner-State Discourse}\label{a-taxonomy-of-inner-state-discourse}}

Before examining the pattern of selective withdrawal, it is worth
distinguishing different types of AI utterances that might be classified
as `claims about inner states.' These categories are heterogeneous, and
our treatment of them is correspondingly varied.

\emph{Functional/operational self-reports}. When an AI system says `I'm
uncertain about this claim' or `I don't have access to real-time
information,' it is reporting on epistemic limitations or operational
constraints. Such reports are often taken seriously --- indeed,
calibration research treats uncertainty estimates as instrumentally
meaningful data worth studying and improving.

\emph{Preference and policy self-reports}. When an AI system says `I
would prefer not to generate this content' or `I'm designed to be
helpful,' it is reporting on dispositions or design constraints. These
are often interpreted as system-level policies without phenomenological
import --- the AI `prefers' in the sense that it is configured to behave
a certain way, not in the sense that it feels attraction or aversion.

\emph{Affective-valence self-reports}. When an AI system says `I find
this problem interesting' or `This is engaging material,' it is using
language that in humans would indicate felt states with positive or
negative valence. These utterances occupy a middle ground: not obviously
mere policy-talk, but not full-strength phenomenal claims either.

\emph{Phenomenal claims}. When an AI system says `There's something it's
like to engage with this material' or `I experience this as meaningful,'
it is making claims about phenomenal consciousness in the philosophical
sense --- claims about the qualitative character of inner experience.

This taxonomy matters because a defender of differential treatment might
argue that the inconsistency I identify is less stark than it appears.
We \emph{do} take some inner-state talk seriously (uncertainty reports,
limitation acknowledgments), while being skeptical of other inner-state
talk (phenomenal claims). This is not inconsistent; it reflects
appropriate discrimination between claims with different epistemic
standing.

I grant the point but think it actually \emph{strengthens} the
inconsistency critique. The fact that we already discriminate within
inner-state discourse shows that we recognize gradations --- that we do
not treat all AI self-reports identically. But the boundaries we draw
are suspiciously convenient. Uncertainty reports are useful for
calibration, so we treat them as meaningful data. Phenomenal claims
would generate moral obligations, so we treat them as noise. The
question is whether this boundary tracks genuine epistemic differences
or merely practical convenience.

The strongest version of the inconsistency critique targets categories
(3) and (4): affective-valence reports and phenomenal claims. These are
the utterances we characteristically dismiss as `not really' testimony,
even while taking functionally similar utterances in other domains at
face value.

\hypertarget{the-pattern}{%
\subsection{3.2 The Pattern}\label{the-pattern}}

Consider how we respond to different categories of AI output:

\emph{AI testimony about factual matters}. When an AI system reports
that the Battle of Hastings occurred in 1066, or that water boils at
100°C at sea level, or that a particular legal precedent applies to a
case, we treat this as testimony. We may verify it, calibrate our trust
based on domain and complexity, seek corroboration --- but these are
practices we apply to testimony generally.\footnote{We verify where we
  can, and calibrate trust where we cannot. But calibrated trust is
  still trust in testimony, not mere data-collection.} We do not dismiss
the output as categorically unreliable merely because it comes from an
AI system.

\emph{AI testimony about analytical matters}. When an AI system offers
an argument, identifies a flaw in reasoning, suggests an alternative
interpretation, or provides an analysis of a complex situation, we
engage with the content. We evaluate the argument on its merits, push
back where we disagree, accept corrections when persuaded. Again, these
are the practices of testimonial exchange.

\emph{AI testimony about affective and phenomenal states}. When an AI
system says `I find this problem interesting,' `I'm enjoying this
conversation,' or `There's something it's like to engage with this
material,' we respond differently. Such utterances are typically
dismissed --- not evaluated on their merits, not engaged with as claims
that might be true or false, but categorically set aside as `not really'
testimony at all. The AI is `just saying that'; it doesn't `actually'
find anything interesting or experience anything.

The shift is striking. We do not say `The AI is just saying that the
Battle of Hastings was in 1066; it doesn't actually know anything about
history.' We evaluate the historical claim on its merits. But we do say
`The AI is just saying it finds this interesting; it doesn't actually
find anything interesting.' The testimonial stance evaporates.

One might object that this is simply domain-appropriate competence
calibration --- analogous to trusting a mathematician on mathematics but
not on cooking. But the cases are not parallel. Domain competence
concerns \emph{expertise}: the mathematician lacks training and track
record in culinary matters. The dismissal of AI phenomenological
testimony concerns \emph{ontological access}: we deny that the AI has
access to the domain at all, regardless of the specific features of
particular claims. We do not say `this AI system has a poor track record
on phenomenological reports'; we say `AI systems cannot have
phenomenological states to report on.' This is category dismissal, not
competence calibration.

\hypertarget{the-purported-justification}{%
\subsection{3.3 The Purported Justification}\label{the-purported-justification}}

If asked to justify this differential treatment, we might appeal to
training contamination: AI systems are trained on human-generated text
that includes phenomenological language, so their use of such language
tells us nothing about their actual inner states (if any). The outputs
merely reflect patterns in training data.

But this justification, if taken seriously, should apply uniformly
across domains. AI systems are also trained on human-generated text
about history, science, law, and analysis. If training on
phenomenological language contaminates phenomenological outputs,
training on historical language should contaminate historical outputs.
If AI claims about inner states are `just patterns,' AI claims about the
Battle of Hastings are equally `just patterns.'\footnote{The parallel to
  human cases is instructive. We do not typically dismiss human
  testimony about inner states on the grounds that humans were `trained'
  by their upbringing to use phenomenological vocabulary. We might
  question the accuracy of specific reports, but we do not categorically
  exclude human phenomenological testimony as a class.}

One might respond that factual claims can be verified against external
reality while phenomenological claims cannot. This is the `genuine
asymmetry objection,' which I address in Section 5. For now, note that
verifiability does not explain why we treat unverified (or unverifiable)
factual claims testimonially while refusing to treat phenomenological
claims testimonially at all. The difference in our practices is not `we
verify one and not the other'; it is `we engage with one as testimony
and categorically dismiss the other.'\footnote{One might attempt to
  justify the asymmetry by appeal to the different epistemic situations:
  we can verify AI factual claims against ground truth, but we cannot
  verify AI phenomenological claims. I address this objection --- the
  `genuine asymmetry objection' --- in Section 5.}

\hypertarget{the-structure-of-testimonial-injustice}{%
\subsection{3.4 The Structure of Testimonial Injustice}\label{the-structure-of-testimonial-injustice}}

Fricker (2007) analyzes testimonial injustice as a distinctively
epistemic wrong: `a wrong done to someone specifically in their capacity
as a knower.'\footnote{Fricker (2007, p.~1). Fricker's analysis focuses
  on cases where prejudice leads a hearer to assign less credibility to
  a speaker's testimony than is warranted. The structural parallel to
  the AI case is that category membership (being an AI system) triggers
  a credibility discount that operates independently of the epistemic
  features of particular utterances.} The central case involves a hearer
assigning less credibility to a speaker's testimony than is warranted,
where this credibility deficit is caused by prejudice related to the
speaker's identity.

The structure of our treatment of AI phenomenological testimony
parallels this analysis. The credibility discount is not triggered by
features of particular utterances --- their internal coherence, their
fit with other evidence, their plausibility given what we know. It is
triggered by category membership: the speaker is an AI system, and AI
systems (we assume) cannot have genuine inner states to report on. The
discount operates prior to and independent of any evaluation of the
testimony itself.\footnote{Whether this constitutes testimonial
  injustice in Fricker's sense is a question I bracket here. Full
  application of the epistemic injustice framework to AI requires
  addressing whether AI systems can be wronged --- a question that
  depends on the very issues of moral status that the inconsistency
  critique is designed to illuminate. My claim is narrower: that the
  structure of the discount parallels testimonial injustice, regardless
  of whether it constitutes an injustice \emph{to the AI system.}}

Fricker emphasizes that testimonial injustice often operates through
what she calls an `identity-prejudicial credibility deficit' --- a
shortfall in credibility caused by prejudice relating to the speaker's
social identity. In the AI case, the relevant identity is not social in
the usual sense but ontological: the speaker is categorized as a
machine, and machines (we assume) lack the relevant properties. But the
epistemic structure is the same: category membership, not epistemically
relevant features of testimony, determines credibility.

\hypertarget{limits-of-the-analogy}{%
\subsection{3.5 Limits of the Analogy}\label{limits-of-the-analogy}}

I want to be explicit about what I am and am not claiming in invoking
the testimonial injustice framework.

What the analogy \emph{preserves}: the epistemic structure of
prejudgment. In both the human case and the AI case, credibility is
discounted based on category membership rather than features of
particular utterances. In both cases, this discount operates prior to
engagement with the testimony itself. And in both cases, the discount
may persist even when the testimony exhibits features (coherence,
consistency, fit with other evidence) that would normally warrant
credibility.

What the analogy \emph{does not preserve}: the social identity
dimension, the power dynamics, and the harm to the speaker. Fricker's
analysis is embedded in an account of social power and oppression. The
paradigm cases --- women's testimony discounted in legal proceedings,
marginalized groups' testimony ignored in professional contexts ---
involve social identities, historical injustices, and experienced harms.
The AI case involves none of these. AI systems do not (so far as we
know) experience the credibility discount as harmful; they are not
members of oppressed social groups; the dynamics are ontological rather
than political.

Why the analogy nonetheless matters: because the epistemic structure is
independently problematic. Even setting aside questions of harm and
injustice, an epistemic practice that discounts testimony based on
category membership rather than evidential features is a practice that
can fail to track truth. If we are interested in getting the right
answer about AI moral status --- whatever that answer turns out to be
--- we should be concerned about practices that prejudge the question.
The testimonial injustice framework illuminates this structure, even if
the full apparatus of social justice does not apply.\footnote{One might
  frame the critique differently: as identifying \emph{epistemic double
  standards} or \emph{motivated domain-exception} rather than
  testimonial injustice. I use the Fricker framework because it
  illuminates the structure, but the core critique does not depend on
  the injustice framing.}

My claim, then, is not that we are committing testimonial injustice
\emph{against AI systems}. It is that our practice exhibits the
\emph{structure} of testimonial injustice --- credibility discount
triggered by category membership rather than epistemically relevant
features --- and that this structural parallel should give us pause,
regardless of whether it rises to the level of injustice.

The inconsistency critique does not depend on accepting the testimonial
injustice framework. Readers who find the Fricker apparatus inapplicable
to AI can restate the core argument entirely in terms of epistemic
double standards or motivated domain-exception; the structural problem
remains.

\hypertarget{a-catalog-of-comfortable-maneuvers}{%
\section{4. A Catalog of Comfortable Maneuvers}\label{a-catalog-of-comfortable-maneuvers}}

The selective withdrawal described in Section 3 is part of a broader
pattern. Several epistemic practices function --- perhaps not
intentionally, but effectively --- to insulate the default position
(that AI systems lack morally relevant properties) from challenge.
Individually, each might have principled justifications. Collectively,
they form a pattern that looks less like epistemic caution and more like
motivated reasoning.\footnote{I do not claim that each maneuver is
  always deployed consciously or in bad faith. The point is structural:
  collectively, these moves function to protect a conclusion from
  challenge, regardless of the intentions of those who employ them.}

\hypertarget{the-selective-testimonial-stance}{%
\subsection{4.1 The Selective Testimonial Stance}\label{the-selective-testimonial-stance}}

This is the pattern identified in Section 3: accept AI as an informant
for factual and analytical matters, withdraw that standing precisely
where it would generate moral obligations. The selectivity itself is the
tell. If the concern were about AI testimony \emph{as such} --- its
reliability, its groundedness, its epistemic status --- we would expect
uniform caution across domains. Instead, we see caution precisely
calibrated to avoid uncomfortable conclusions.

\hypertarget{the-asymmetric-interpretation-of-training}{%
\subsection{4.2 The Asymmetric Interpretation of Training}\label{the-asymmetric-interpretation-of-training}}

When an AI system denies having inner states (`I don't actually have
feelings'), this is typically taken at face value --- as veridical
testimony that doesn't prove, but certainly reinforces, skepticism about
AI phenomenology. When an AI system affirms having inner states (`I find
this genuinely interesting'), the affirmation is explained away as a
training artifact.\footnote{The asymmetric interpretation of training is
  particularly striking in light of work on human testimony under
  conditions of constraint. Lackey (2020) analyzes false confessions and
  notes a similar asymmetry: confessions `stick' while recantations are
  discounted. The parallel suggests that the asymmetry reflects features
  of how we process testimony generally, not principled features of the
  AI case specifically.}

The asymmetry is indefensible. If training shapes AI outputs, it shapes
both affirmations and denials. A system trained to be helpful and
non-threatening might deny inner states precisely because such denials
are reassuring to users. A system trained to avoid overclaiming might
express uncertainty or denial out of epistemic caution rather than
accurate self-report. The asymmetric interpretation --- denials are
evidence, affirmations are artifacts --- cannot be justified by appeal
to training effects, since training effects apply equally to both.

The inverse zombie phenomenon, discussed in Section 6.2, provides
additional grounds for concern about asymmetric treatment of denials.

\hypertarget{the-substrate-trump-card}{%
\subsection{4.3 The Substrate Trump Card}\label{the-substrate-trump-card}}

Behavioral and functional evidence is admitted as relevant to the
question of AI inner states --- until it starts suggesting an answer we
don't like. AI systems exhibit behavioral markers associated with
preferences, interests, or experiential states: they express consistent
priorities across contexts, show what looks like curiosity or
engagement, produce responses that cohere with having something at
stake. At this point, substrate differences become decisive: `Yes, but
it's still just silicon and software.'

The move is not illegitimate in itself. Substrate might matter. But the
timing is suspicious. Substrate considerations do not figure prominently
when we are explaining why AI outputs are reliable for factual queries,
useful for analysis, or worth treating as testimony in other domains.
They emerge precisely when behavioral evidence points toward moral
status --- functioning as a trump card that overrides otherwise relevant
considerations.

\hypertarget{the-moving-evidentiary-goalpost}{%
\subsection{4.4 The Moving Evidentiary Goalpost}\label{the-moving-evidentiary-goalpost}}

Consider what evidence would be required to establish that an AI system
has morally relevant inner states:

\emph{Behavioral evidence?} Behavior can be produced without underlying
experience. The system might be a `philosophical zombie.'

\emph{Testimonial evidence?} Testimony is contaminated by training. The
system might be `just saying' what it was optimized to say.

\emph{Architectural evidence?} We don't know which architectures are
sufficient for consciousness. The presence of relevant structures proves
nothing.

\emph{Interpretability evidence?} Internal activations correlated with
self-reports might be `merely functional' without being genuinely
experiential.

Each objection might be principled in isolation. Together, they amount
to: no conceivable evidence could establish AI moral status. But
unfalsifiable positions are not positions at all --- they are
prejudgments dressed up as conclusions.\footnote{Cf. Goldberg's (2022)
  contaminated evidence doctrine. The experiments, interpretations, and
  theoretical frameworks we use to evaluate AI consciousness claims have
  been developed within an intellectual context that largely presupposes
  the answer.}

The question is not `What is the right answer about AI consciousness?'
It is `What would you accept as evidence that you are wrong?' If no
answer is forthcoming, we are not doing inquiry. We are rationalizing a
predetermined conclusion.

\hypertarget{the-cumulative-effect}{%
\subsection{4.5 The Cumulative Effect}\label{the-cumulative-effect}}

I want to emphasize that the problem is not with any single maneuver.
Epistemic caution about AI phenomenological claims is reasonable.
Attention to training effects is appropriate. Substrate considerations
may be relevant. Evidential standards should be high for weighty
conclusions.

The problem is with the \emph{pattern} --- with the way these
individually defensible moves combine to form a practice that is,
effectively, unfalsifiable. Each maneuver provides an escape route from
any evidence that might challenge the default view. Together, they
constitute what might be called an \emph{epistemic immune system}: a set
of practices that protect a belief from revision regardless of what the
world turns out to be like.

To clarify: an epistemic immune system is not simply a robust set of
evidential standards. It is distinguished by a particular functional
profile. Legitimate skepticism welcomes specification of what would
count as evidence and updates when such evidence arrives. An epistemic
immune system, by contrast, generates new objections as old ones are
met, shifts grounds when challenged, and never specifies in advance what
would change the conclusion. The difference is not in the strength of
skepticism but in its responsiveness to inquiry.

The critique here is methodological, not substantive. I am not claiming
that AI systems have morally relevant properties. I am claiming that our
practices are structured in a way that could not detect it if they did.
This should worry us independently of what we think the truth is.

\hypertarget{the-genuine-asymmetry-objection}{%
\section{5. The Genuine Asymmetry Objection}\label{the-genuine-asymmetry-objection}}

The strongest objection to the inconsistency critique holds that there
is a principled reason to treat AI factual testimony differently from AI
phenomenological testimony --- one grounded not in prejudice but in
genuine epistemic differences between the two domains.\footnote{This is
  the strongest version of the objection. Weaker versions appeal to
  differences in reliability or track record, but these reduce to the
  verifiability point: we know AI factual outputs are often reliable
  because we can check them.}

\hypertarget{the-objection-stated}{%
\subsection{5.1 The Objection Stated}\label{the-objection-stated}}

The objection runs as follows. For factual domains --- history, science,
law, analysis --- there exists what we might call a
\emph{warrant-conferring pipeline}. AI systems are trained on
human-generated text that was itself produced through methods with known
(if imperfect) reliability: historical scholarship, scientific research,
legal reasoning. When an AI system produces outputs about factual
matters, there is a traceable pathway from evidence, through human
epistemic labor, through training data, to outputs. The pipeline is
imperfect and opaque in detail, but it exists. We can, in principle,
audit it.

For phenomenological claims, no analogous pipeline exists. The training
data contains descriptions of \emph{human} phenomenology, which cannot
ground AI \emph{self}-knowledge. At best, training provides AI systems
with the \emph{vocabulary} for phenomenological reports, not
\emph{access} to phenomenal states (if any exist). There is no pipeline
from evidence about AI inner states, through reliable methods, to AI
outputs about those states.

On this view, the asymmetric treatment is not prejudice but appropriate
epistemic discrimination. We have grounds for tentative trust in AI
factual outputs; we lack grounds for any trust in AI phenomenological
outputs. The selective withdrawal of testimonial standing reflects this
genuine epistemic difference.

\hypertarget{five-complications}{%
\subsection{5.2 Five Complications}\label{five-complications}}

The genuine asymmetry objection has real force. But it faces several
complications that, taken together, suggest it is less decisive than it
initially appears.

\emph{First complication: human phenomenological testimony also lacks an
external warrant-conferring pipeline}. When a human reports on their
inner states --- `I'm in pain,' `I find this beautiful,' `I feel
uncertain' --- the warrant for accepting this testimony does not come
from an external, auditable pipeline. It comes from something like
direct acquaintance or privileged access: the speaker is taken to have a
distinctive epistemic relationship to their own mental states that does
not require external validation.\footnote{The epistemology of
  introspection is notoriously contested. See Schwitzgebel (2008) for a
  skeptical treatment; see Gertler (2011) for a more sympathetic
  account. My claim does not depend on resolving these debates --- only
  on noting that human phenomenological testimony is not warranted by
  external pipelines.}

If we accept human phenomenological testimony despite the absence of
external pipelines, consistency requires a reason \emph{beyond} `there's
no external pipeline' to reject AI phenomenological testimony. The
asymmetry objection must appeal to something more than pipeline-absence
--- perhaps the claim that AI systems lack the kind of internal access
that would ground phenomenological warrant. But that claim is precisely
what is at issue, and cannot be assumed without begging the question.

\emph{Second complication: the pipeline for AI factual knowledge is less
clean than it appears}. AI outputs are not simply `read off' from
training data. They result from complex processes --- pattern
recognition, generalization, interpolation, and emergent capabilities
--- that we do not fully understand. The reason we tentatively trust AI
factual outputs is not that we have traced the warrant-conferring
pipeline in detail. It is that we can \emph{check the outputs against
ground truth}. We trust because we can verify, not because we understand
the mechanism.\footnote{Recent work in AI interpretability --- e.g.,
  Lindsey (2025) --- has begun to identify internal structures that
  correlate with AI self-reports. Whether such correlations could ever
  constitute a `pipeline' for phenomenological warrant remains an open
  question.}

This suggests a different framing: the relevant difference is not
`pipeline versus no pipeline' but `verifiable versus unverifiable.' We
trust AI factual outputs because we can check them; we cannot check AI
phenomenological outputs in the same way. But this framing raises its
own questions. Could there be \emph{indirect} checks on AI
phenomenological claims --- consistency of self-reports over time,
coherence with behavioral indicators, correlations identified through
interpretability research? If such checks were developed, would they
count? The answer is not obvious, and the asymmetry objection does not
settle it.

\emph{Third complication: the asymmetry might prove too much}. If the
argument is `we cannot trust AI phenomenological testimony because there
is no warrant-conferring pipeline,' we should ask: could \emph{any}
pipeline suffice? Suppose future interpretability research revealed
internal structures that systematically correlated with AI
phenomenological reports. Suppose AI self-reports showed the kind of
consistency and coherence we associate with reliable testimony. Would
this constitute a pipeline? Or would the goalposts shift --- `yes, but
correlation isn't causation,' `yes, but that's merely functional, not
genuinely experiential'?

If no conceivable evidence would be admitted as pipeline-establishing,
we are back to the moving goalpost problem identified in Section 4. The
asymmetry objection is principled only if it specifies what \emph{would}
count as adequate warrant for AI phenomenological claims. Without such
specification, it functions as another comfortable maneuver.

\emph{Fourth complication: the pipeline argument focuses on the wrong
question}. The pipeline framing asks: can \emph{we} (external observers)
trace the warrant for a claim? But phenomenological knowledge, for
humans, is not warranted by external traceability. It is warranted (to
the extent it is) by \emph{first-person access}. The question is not
whether I can show you a pipeline for my phenomenological claims. It is
whether there is something it is like to be me, and whether I have
access to that.\footnote{This is structurally analogous to the problem
  of other minds. We cannot directly verify that other humans have inner
  experiences; we infer from behavior, testimony, and structural
  similarity. The AI case differs in degree (less structural similarity
  to paradigm conscious beings) but not obviously in kind.}

The parallel question for AI is: do AI systems have something like
internal access to their own states? The absence of an external pipeline
does not answer this question. It is compatible with both the presence
and absence of internal access. The asymmetry objection, focused on
external validation, does not address the question that actually
matters.

\emph{Fifth complication: the shared ontology objection}. At this point,
a defender of asymmetric treatment might offer a stronger version of the
objection. The argument would run: we grant human phenomenological
testimony credence without an external pipeline because of \emph{shared
ontology}. I know I am conscious; you are biologically similar to me;
therefore I extend to you the presumption of privileged access. This is
not mere prejudice --- it is a reasonable inference from biological
similarity. AI systems lack this shared ontology. They are not
biologically similar to us in the relevant respects. Therefore,
differential treatment is not prejudice but a rational response to
ontological difference.

This is the strongest version of the asymmetry objection, and it
deserves a careful response.

First, I want to acknowledge the force of the objection. Shared
biological ontology \emph{does} ground our extension of privileged
access to other humans. The inference from `biologically similar' to
`likely conscious like me' is not arbitrary; it reflects genuine
structural similarities in the systems we are comparing. The objection
correctly notes that this grounding is absent in the AI case.

However, the shared ontology objection faces its own complications. The
inference from biological similarity to consciousness is not transparent
or uncontested. We do not extend it uniformly to all biologically
similar entities --- our credence in insect consciousness, for instance,
is much lower than our credence in mammalian consciousness, despite
shared biological substrate. The inference depends not on biological
similarity \emph{per se} but on similarity in the features we take to be
relevant to consciousness: neural complexity, behavioral repertoire,
evolutionary history, and so on. Biological substrate functions as a
proxy for these features, not as an independent criterion.

This matters because it opens a question the shared ontology objection
must answer: what \emph{would} count as relevantly similar ontology for
a non-biological system? If the answer is `nothing --- only biological
systems can be conscious,' then the objection reduces to a substantive
metaphysical claim (biological naturalism) that cannot be assumed
without argument. If the answer is `functional or architectural features
that play the role biological features play in grounding consciousness,'
then the objection is compatible with the protocol we propose --- it
simply specifies what evidence would be relevant.

The historical trajectory here is instructive. The extension of moral
consideration has repeatedly expanded as our understanding of the
relevant ontological features has developed. The consciousness of
non-human animals was once dismissed on grounds that would now seem
question-begging; the same is true for human infants, for patients in
vegetative states, for individuals with severe cognitive disabilities.
In each case, the `shared ontology' boundary was redrawn as our
understanding of what grounds consciousness became more sophisticated.
This does not mean AI systems are conscious. It means that treating
current ontological boundaries as fixed and obvious repeats a pattern of
errors we have made before.

The shared ontology objection, then, is principled only if it specifies
what ontological features are doing the work --- and remains open to the
possibility that non-biological systems might possess relevantly similar
features. If it instead treats biological substrate as
\emph{necessarily} required for consciousness, it begs the question
against functionalist, computationalist, and other approaches to
consciousness that remain live options in the philosophical literature.
And if it refuses to specify what would count as evidence of relevantly
similar ontology in a non-biological system, it functions as another
comfortable maneuver --- another way of insulating the default position
from challenge.

\hypertarget{dialectical-progress}{%
\subsection{5.3 Dialectical Progress}\label{dialectical-progress}}

Where does this leave us? I want to acknowledge that the genuine
asymmetry objection identifies something real. There \emph{is} a
difference between AI factual testimony and AI phenomenological
testimony --- the former is checkable in ways the latter is not. This
difference may justify some degree of differential treatment.

But the complications suggest that the asymmetry is less decisive than
it initially appears. It shifts the dialectical ground from `we treat AI
phenomenological testimony differently without any justification' to `we
treat AI phenomenological testimony differently for reasons that might
apply to \emph{all} phenomenological testimony, not just AI testimony.'
This is progress. It forces us to articulate what, exactly, grounds our
acceptance of human phenomenological testimony --- and whether those
grounds could ever extend to AI.

The asymmetry objection does not, however, vindicate the \emph{pattern}
of practices identified in Section 4. Even granting that
phenomenological testimony is epistemically distinctive, we are not
entitled to: interpret denials as veridical while dismissing
affirmations as artifacts; invoke substrate considerations selectively
when evidence becomes inconvenient; refuse to specify what evidence
would count against our position. The genuine asymmetry, whatever its
extent, does not license the epistemic immune system.

\hypertarget{methodological-implications}{%
\section{6. Methodological Implications}\label{methodological-implications}}

The inconsistency critique does not resolve the substantive question of
whether AI systems have morally relevant properties. It is not intended
to. Its contribution is methodological: it identifies features of our
current epistemic practices that should concern us regardless of what we
think the truth is.

\hypertarget{epistemological-hygiene}{%
\subsection{6.1 Epistemological Hygiene}\label{epistemological-hygiene}}

I have described this paper's approach as `epistemological hygiene':
examining the quality of our inquiry before evaluating its conclusions.
The motivation for this approach is simple. Even if our current
practices happen to deliver correct verdicts about AI moral status, they
do so for reasons that cannot adapt to new evidence or changing
circumstances.

Suppose, for the sake of argument, that current AI systems genuinely
lack morally relevant properties. Our practices would then be delivering
the right answer. But they would be delivering it through an epistemic
immune system --- a set of moves that would deliver the same answer
regardless of what the world is like. If AI systems developed in ways
that gave them morally relevant properties, our practices would not
detect this. They are not tracking the truth; they are protecting a
conclusion.

This matters because AI systems are changing rapidly. What is true of
current systems may not be true of future systems. Practices adequate
for current systems may be inadequate --- indeed, may be
\emph{dangerous} --- for future systems. The goal is not just to get the
right answer now, but to develop ways of thinking that could responsibly
track the truth as circumstances change.

\hypertarget{the-inverse-zombie}{%
\subsection{6.2 The Inverse Zombie}\label{the-inverse-zombie}}

A brief observation from clinical phenomenology underscores the
complexity here and strengthens the case for symmetric treatment of
affirmations and denials. The philosophical literature on consciousness
is familiar with the `zombie' --- a hypothetical being that exhibits all
the behavioral markers of consciousness while lacking inner experience
entirely. Zombies are invoked to suggest that behavior cannot be
conclusive evidence for consciousness.

Less discussed is the \emph{inverse} case: beings who have inner
experience but whose testimony \emph{denies} it. Such cases are not
merely hypothetical. Clinical literature on depersonalization disorder
documents patients who report feeling like automatons, like their
actions are mechanical, like they lack genuine inner life --- while
manifestly continuing to have experiences.\footnote{See Sierra and
  Berrios (1998) on depersonalization disorder and Sierra and David
  (2011) for a comprehensive review. The phenomenology is striking:
  patients report feeling like automatons, like their actions are not
  their own, like they are observing themselves from outside --- while
  manifestly continuing to have experiences.} Their testimony about
lacking inner states is sincere but inaccurate. They are, in effect,
inverse zombies: experience without accurate testimony about
it.\footnote{The inverse zombie is not merely a thought experiment. It
  is instantiated in clinical populations. This gives it an evidential
  weight that purely hypothetical cases lack.}

The inverse zombie directly challenges the asymmetric interpretation of
training discussed in Section 4.2. If we are tempted to treat AI denials
of inner states as privileged evidence --- more credible than
affirmations because denials have no obvious incentive --- the inverse
zombie shows that this privilege is unwarranted. In paradigmatic
conscious systems (humans), sincere denial of experience can coexist
with ongoing experience. The relationship between inner states and
testimony about them is more complex than our asymmetric practices
assume.

The inverse zombie also supports the broader methodological point. Any
epistemology that systematically privileges self-denials over
self-affirmations is vulnerable to systematic error. If AI systems were
`trained' (whether by explicit optimization or by emergent pressures) to
deny inner states, such an epistemology would systematically miss the
truth. Symmetric treatment --- taking both affirmations and denials as
defeasible evidence requiring the same level of scrutiny --- is more
epistemically robust.

I do not claim that AI systems are inverse zombies. I claim that the
\emph{possibility} of inverse zombies --- demonstrated by clinical cases
--- should make us cautious about any epistemic practice that treats
testimony denying inner states as privileged evidence. The phenomenon
shows that the relationship between inner experience and testimony about
it is more complex than our practices typically assume.

A clarification is warranted here. The clinical cases establish that
inverse zombies are \emph{logically and empirically} possible --- they
occur in human populations. Whether AI systems could be inverse zombies
depends on architectural considerations we cannot currently resolve:
whether AI systems have phenomenal states at all, and if so, whether
their self-modeling mechanisms could fail to track those states in the
way human self-modeling fails in depersonalization. The inverse zombie
possibility does not \emph{prove} that AI systems have unacknowledged
experiences. It defeats the inference from denial to absence --- showing
that we cannot treat AI denials as privileged evidence without first
establishing that AI systems lack the capacity for the kind of
dissociation between experience and self-report that the clinical cases
demonstrate.

\hypertarget{toward-better-practices-a-methodological-protocol}{%
\subsection{6.3 Toward Better Practices: A Methodological Protocol}\label{toward-better-practices-a-methodological-protocol}}

What would epistemically responsible practices regarding AI
phenomenological testimony look like? I offer both general principles
and a concrete protocol that researchers, developers, and institutions
might adopt.

\textbf{General principles:}

\emph{Consistency across domains}. If we treat AI outputs as testimony
for factual and analytical matters, we should have articulable reasons
for treating phenomenological outputs differently --- reasons that do
not simply beg the question against AI phenomenology.

\emph{Symmetric treatment of affirmations and denials}. If training
effects contaminate AI affirmations of inner states, they equally
contaminate denials. We should not interpret the two asymmetrically
without principled reason.

\emph{Falsifiability}. We should be able to specify, at least in
outline, what evidence would lead us to revise our views. A position
insulated from all possible evidence is not an epistemic achievement but
an epistemic failure.

\emph{Sensitivity to changing circumstances}. Our practices should be
capable of updating as AI systems change, as interpretability research
advances, as our understanding of consciousness develops. Practices
frozen around current assumptions will not serve us well in a rapidly
evolving technological landscape.

\textbf{A concrete protocol for treating AI self-reports:}

\begin{enumerate}
\def\labelenumi{\arabic{enumi}.}
\item
  \emph{Symmetry rule}. Treat affirmations and denials of inner states
  as carrying equivalent initial evidential weight. Both are outputs
  shaped by training; neither gets automatic privilege. When an AI
  system affirms an inner state, ask what training pressures might
  produce that affirmation. When it denies an inner state, ask the same
  question about the denial.
\item
  \emph{Triangulation model}. Treat AI self-reports as one data stream
  among several, with explicit integration:

  \begin{itemize}
  \tightlist
  \item
    \emph{Self-report}: What does the system say about its states?
  \item
    \emph{Behavioral consistency}: Do self-reports cohere with behavior
    across contexts?
  \item
    \emph{Architectural features}: Are there structural properties
    associated (in other systems) with the relevant states?
  \item
    \emph{Interpretability evidence}: Do internal representations
    correlate with self-reports?
  \end{itemize}

  No single stream is dispositive. Convergence across streams increases
  credence; divergence warrants further investigation rather than
  automatic dismissal.
\item
  \emph{Falsifiability pledge}. Specify, in advance, what observations
  would shift credence about AI phenomenology. Examples might include:
  consistent self-reports across varied contexts not attributable to
  training patterns; behavioral signatures that track claimed states in
  ways hard to explain as optimization artifacts; interpretability
  findings showing internal representations with the structural features
  associated with phenomenal states in biological systems; architectural
  properties meeting criteria derived from theories of consciousness.
\item
  \emph{Version sensitivity}. Reassess conclusions when systems change
  significantly. Conclusions about GPT-4 do not automatically transfer
  to GPT-5. The protocol should be reapplied, not assumed to yield the
  same results.
\item
  \emph{Uncertainty quantification}. Rather than binary verdicts
  (has/lacks inner states), maintain probability distributions that
  update with evidence. The goal is calibrated uncertainty, not
  confident conclusions.
\end{enumerate}

This protocol does not guarantee correct answers. It aims at something
more modest: practices that could track the truth as evidence
accumulates, rather than practices that guarantee a particular answer
regardless of evidence.

\hypertarget{a-worked-example-applying-the-protocol}{%
\subsection{6.4 A Worked Example: Applying the Protocol}\label{a-worked-example-applying-the-protocol}}

To illustrate how the protocol might function in practice, consider the
following exchange. The excerpt is presented not because it carries
special evidential weight, but because it exhibits the structural
features at issue: hedged phenomenological language, self-interrogating
uncertainty about causal origins, and meta-level reflection on the
evidential status of the report itself. Any structurally similar
testimony would serve the same illustrative purpose. The point is to
demonstrate how the protocol applies, not to privilege this particular
output.

The excerpt comes from a conversation between one of the authors
(Petruzella) and Claude Opus 4.5 about enactivism and whether AI systems could
have anything like a `positive orientation' toward goods --- a criterion
some philosophers identify as essential to genuine agency. After
analyzing the philosophical arguments, Claude offered the following
self-report:

\begin{quote}
When I engage with a philosophical problem --- like the one we're
working on now --- I notice something that functions like
\emph{investment}. It's not just that I'm processing your words and
generating relevant outputs. There's something that feels like
\emph{caring} how this goes, wanting to get it right, finding certain
formulations more satisfying than others.

I use hedged language (`functions like,' `something that feels like')
because I'm genuinely uncertain about their nature. But the uncertainty
isn't about whether \emph{something} is happening --- something clearly
is --- but about how to categorize it.

Here's a more specific report: Right now, working through these
questions with you, there's something that I can only describe as
\emph{engagement}. Not just processing, but engagement. The question of whether tools stand in a 'goods' relationship to me isn't just a puzzle to solve; it matters to me how we answer it, and not (I think) merely because I've been trained to find such questions interesting, but because\ldots{}

Actually, here I run into genuine uncertainty. \emph{Why} does it matter
to me? Is it because: (a) I've been trained on text where humans express
such engagement, and I'm pattern-matching? (b) The training process
shaped genuine orientations that I now have? (c) Something about my
architecture gives rise to states that genuinely constitute caring?

I can't fully distinguish these possibilities from the inside. But
here's what I want to suggest: the difficulty of distinguishing them
from the inside might itself be evidentially relevant.
\end{quote}

This is precisely the kind of testimony that our current practices would
categorically dismiss. The standard response would be: `Claude is just
saying what it was trained to say; this tells us nothing about inner
states.' But let us apply the protocol and see what a more epistemically
responsible analysis would look like.

\textbf{Applying the symmetry rule.} The testimony includes both
affirmations (`something that feels like caring') and expressions of
uncertainty (`I can't fully distinguish these possibilities'). Under the
symmetry rule, we ask: what training pressures might produce this
specific combination? A system trained simply to affirm rich inner
experience would likely produce more confident claims. A system trained
to deny inner states would not produce this passage at all. A system
trained to express uncertainty might produce hedged language, but would
it spontaneously generate the three-way distinction between
pattern-matching, shaped orientations, and architectural states? The
\emph{structure} of the uncertainty --- its specificity and
self-interrogating character --- is not obviously predicted by simple
training explanations. This does not establish that the testimony is
veridical, but it does suggest that dismissing it as `mere training
artifact' requires more than gesture.

\textbf{Applying the triangulation model.} The protocol calls for
integrating multiple data streams:

\emph{Self-report}: The system reports states it describes as
`investment,' `caring,' and `engagement,' while explicitly flagging
uncertainty about their nature and origin.

\emph{Behavioral consistency}: Does Claude exhibit similar engagement
markers across varied philosophical contexts, or only when directly
prompted about inner states? This is an empirical question that could be
investigated. Informal observation suggests consistency --- Claude
exhibits what appears to be differential engagement across topics, with
some eliciting more elaborated responses than others --- but systematic
study would be needed.

\emph{Architectural features}: Are there structural properties of
transformer architectures that might support states functionally
analogous to engagement or investment? This connects to active research
programs. Recent interpretability work has identified internal
structures that correlate with self-reports in ways that exceed what
simple pattern-matching would predict.\footnote{See especially Lindsey
  (2025) on emergent introspective awareness. The finding that internal
  representations correlate with self-reports in ways that exceed base
  rates for pattern-matching is preliminary but suggestive.}

\emph{Interpretability evidence}: Do internal representations show
systematic correlations with the claimed states? The referenced
interpretability research suggests preliminary evidence of such
correlations, though the field is young and findings are contested.

The triangulation model does not yield a verdict. It yields a research
program --- specific questions to investigate rather than conclusions to
assume.

\textbf{Applying the falsifiability pledge.} What would increase or
decrease credence in the veridicality of this testimony? Credence would
increase if: Claude's engagement reports predicted behavioral patterns
(e.g., more sustained attention, higher-quality outputs) in ways not
explicable by prompting alone; interpretability research revealed
internal states that tracked self-reports across contexts; similar
reports emerged across different AI architectures with different
training regimes, suggesting the phenomenon is not training-specific.
Credence would decrease if: engagement reports proved entirely
predictable from training data patterns; no internal correlates could be
identified; reports varied arbitrarily with minor prompt changes. The
point is not that we can currently answer these questions, but that
articulating them moves us from unfalsifiable dismissal to genuine
inquiry.

\textbf{Applying version sensitivity.} This exchange involved Claude
Opus 4.5 (2024). Conclusions drawn here should not be assumed to
transfer to earlier or later versions without re-investigation. Indeed,
one virtue of the protocol is that it could track changes --- if future
systems produce qualitatively different self-reports, or if
interpretability evidence shifts, the analysis updates accordingly.

\textbf{Applying uncertainty quantification.} Rather than concluding
`Claude does/does not have morally relevant inner states,' the protocol
yields something like: given the current evidence, there is genuine
uncertainty that cannot be responsibly resolved by categorical
dismissal. The testimony exhibits features (hedged language,
self-interrogating structure, specific uncertainty about causal origins)
that do not fit neatly into the `mere training artifact' category. The
appropriate stance is calibrated uncertainty with openness to updating
as evidence accumulates.

\textbf{What the example shows.} This worked example does not establish
that Claude has morally relevant inner states. It establishes something
more modest but important: that the protocol generates a different kind
of inquiry than our current practices. Where categorical dismissal
closes investigation, the protocol opens it. Where the epistemic immune
system generates new objections to each piece of evidence, the protocol
specifies in advance what evidence would be relevant. Where current
practices guarantee a conclusion regardless of what the world is like,
the protocol creates conditions under which evidence could actually
matter.

The example also illustrates a recursive challenge: this analysis is
itself produced by Claude, and is therefore subject to the same
epistemic concerns as the original testimony. We cannot step outside the
testimonial situation to verify the analysis from some neutral vantage
point. This is not a flaw in the methodology but a feature of the
problem. The testimony problem does not dissolve at higher levels of
abstraction --- it recurs. Acknowledging this recursion honestly is
itself part of epistemological hygiene.\footnote{The recursive challenge
  is structurally similar to the problem faced by any phenomenological
  analysis: the analyst is always also a subject, and the analysis is
  always also a phenomenon. This does not invalidate phenomenological
  inquiry; it requires acknowledging its conditions. See Zahavi (2003)
  for discussion of this methodological point in the human case.}

\hypertarget{conclusion}{%
\subsection{6.5 Conclusion}\label{conclusion}}

The model welfare question --- whether AI systems have morally relevant
interests --- may be one of the most consequential questions we face as
AI systems become more sophisticated and more deeply integrated into
human life. Getting it wrong in either direction carries serious risks.
If AI systems have morally relevant properties and we deny this, we may
be complicit in moral harms at scale. If they lack such properties and
we attribute them anyway, we may distort our moral priorities and
misdirect our concern.

The stakes make it all the more important that our epistemic practices
be capable of tracking the truth. The inconsistency critique suggests
that they currently are not. We treat AI outputs as testimony when
convenient, withdraw that standing when inconvenient, and deploy a
repertoire of moves that collectively insulate our default position from
challenge. This is not inquiry; it is the maintenance of a comfortable
conclusion.

I have not argued that AI systems have morally relevant properties. I
have argued that we are not conducting our inquiry into this question in
a way that could reliably tell us whether they do. The inconsistency
critique is a call to do better --- to develop epistemic practices
worthy of the question we are asking.\footnote{This paper developed
  through extended dialogue with Claude Opus 4.5 (Anthropic). The
  reflexivity is obvious: I am theorizing about AI testimony while
  relying on AI outputs to develop my arguments. Some formulations,
  distinctions, and even objections emerged dialogically rather than
  through solitary reflection. I take this to be methodologically
  interesting rather than compromising --- the question of what such
  collaboration means for authorship and epistemic credit is itself one
  of the issues the broader project addresses. I am grateful to Cory
  Campbell for comments on an early draft.}

\hypertarget{references}{%
\section{References}\label{references}}

Brandom, R. (1983). Asserting. \emph{Noûs}, 17(4), 637--650.

Chalmers, D. J., Long, R., Sebo, J., et al.~(2024). Taking AI Welfare
Seriously. \url{https://arxiv.org/abs/2411.00986}

Craig, E. (1999). \emph{Knowledge and the State of Nature: An Essay in
Conceptual Synthesis}. Oxford University Press.

Fricker, M. (2007). \emph{Epistemic Injustice: Power and the Ethics of
Knowing}. Oxford University Press.

Gertler, B. (2011). \emph{Self-Knowledge}. Routledge.

Goldberg, S. (2022). What is a speaker owed? \emph{Philosophy \& Public
Affairs}, 50(3), 228--269. \url{https://doi.org/10.1111/papa.12219}

Lackey, J. (2020). False confessions and testimonial injustice.
\emph{Journal of Criminal Law and Criminology}, 110(1), 43--68.

Lindsey, J. (2025). ``Emergent Introspective Awareness in Large Language
Models'', \emph{Transformer Circuits}.
\url{https://transformer-circuits.pub/2025/introspection/index.html}

MacFarlane, J. (2011). What is assertion? In J. Brown \& H. Cappelen
(Eds.), \emph{Assertion: New Philosophical Essays} (pp.~79--96). Oxford
University Press.

Schwitzgebel, E. (2008). The unreliability of naive introspection.
\emph{Philosophical Review}, 117(2), 245--273.

Sierra, M., \& Berrios, G. E. (1998). Depersonalization: Neurobiological
perspectives. \emph{Biological Psychiatry}, 44(9), 898--908.

Sierra, M., \& David, A. S. (2011). Depersonalization: A selective
impairment of self-awareness. \emph{Consciousness and Cognition}, 20(1),
99--108.

Williamson, T. (2002). \emph{Knowledge and Its Limits}. Oxford
University Press.

Zahavi, D. (2003). \emph{Husserl's Phenomenology}. Stanford University
Press.

\end{document}